\documentclass[twocolumn,english,showpacs,prl]{revtex4}
\usepackage{amsmath}
\usepackage{amssymb}

\usepackage{dcolumn}
\renewcommand\vec[1]{\boldsymbol{#1}}

\usepackage{babel}

\begin{document}

\title{The unreasonable accuracy of the Jastrow approach in many-body physics}

\author{Niels R. Walet}

\email{Niels.Walet@umist.ac.uk}

\author{R. F. Bishop}

\email{R.F.Bishop@umist.ac.uk}

\affiliation{Department of Physics, UMIST, P.O. Box 88, Manchester M60 1QD, UK}

\date{\today{}}

\begin{abstract}
We investigate in a simple model whether a Jastrow-based approach for a many-body 
system containing two-body interactions can be exact. By comparison with 
recent assertions to the contrary, we find that in general the exact wave 
function is \emph{not} purely two-body correlated.  Nonetheless, we show that the 
energy estimate obtained from the calculations is as accurate as it can 
possibly be, suggesting why Jastrow wave functions are such a good choice.
\end{abstract}


\maketitle

A long-standing and ongoing fundamental problem in quantum physics is the 
search for general methods to provide accurate descriptions of strongly 
interacting systems with many degrees of freedom, starting from a microscopic 
Hamiltonian.  One of the first such broadly based techniques, which is still 
in widespread use, was introduced by Feenberg \cite{Feenberg}. He generalized 
a similar variational approach of Jastrow \cite{Jastrow},  the basic idea of 
which is that in a strongly correlated $N$-body system the 
wave function may be approximated as the product of two parts.  The first is 
a simple {}``uncorrelated'' part which incorporates the correct quantum statistics 
(e.g., a Slater determinant for fermions).  The second describes the correlations and
is a product of pairwise Jastrow factors, which 
depends only on the interparticle separations.

Many, if not most, of the key methods in modern quantum many-body theory 
are based on similar ideas, but often with simple but subtle differences 
which have deep implications, as we elaborate below.  For example, the variational 
Monte Carlo (VMC) method often employs a parametrized Jastrow-correlated wave 
function which is optimized using a statistical approach.  Although the VMC 
method is not intrinsically tied to Jastrow wave functions, their usefulness 
has been proven empirically by the high level of accuracy that can be reached 
with such a simple form of trial wave function in many applications in nuclear 
physics and elsewhere \cite{VMC}.  Many authors in many different fields have 
attempted to understand the surprisingly accurate results obtained by Jastrow 
trial wave functions.  For example, Gaudoin \emph{et al.}\ \cite{Gaudoin} have 
shown how by generalizing the random phase approximation of the 
electron gas to the inhomogeneous case, a Slater-Jastrow trial wave function 
arises rather naturally.  They also showed how the uncorrelated Slater 
determinant contains optimal orbitals which are close both to standard 
Hartree-Fock orbitals and to orbitals obtained from density-functional theory 
within the local density approximation, even though neither of these latter 
approaches includes Jastrow factors \emph{a priori}.

Although the use of such simple Jastrow-correlated many-body wave functions 
has been seen to give a large part of the correlation energy in many 
diverse applications, one would also like to improve the method.  Several 
such refinements have been widely used.  Firstly, one may 
extend the two-body correlations to depend
also on internal quantum numbers such as spin.  This is particularly important 
in cases where the interactions are 
state-dependent, such as in 
nuclear physics.  State-dependence in the two-body correlators turns them 
effectively into operators.  The fact that the parameters of a given particle 
occur in many of the Jastrow factors 
then leads to them not 
commuting between themselves, and hence the 
product of $N(N-1)/2$ 
correlation operators needs to be symmetrized. This considerably complicates the formalism, 
and the Fermi hypernetted chain (FHNC) method \cite{FHNC} is one approach that 
deals with it.  It is based on a cluster expansion of the Jastrow correlations, 
and 
the method 
has proven to be quite accurate.  An alternative to the FHNC method for Hamiltonians 
which include tensor and other spin-dependent interactions has been developed by 
Schmidt and Fantoni \cite{FS}, using a purely numerical approach.  
A second extension of the simplest 
Jastrow scheme is the inclusion of the product of all 
3-body scalar cluster correlation functions, $\prod_{i<j<k}f(r_{ij},r_{jk},r_{ki})$, 
and similar products of $n$-body cluster correlation functions with $n>3$, 
as well as the Jastrow product of all 2-body correlation functions, 
$\prod_{i<j}f(r_{ij})$.  Thirdly, the more general method of correlated 
basis functions (CBF) \cite{Feenberg,FHNC,CBF}, which employs a correlated 
basis rather than a single trial wave function as above, is one of the two 
most successful universal many-body methods available today.  The correlation 
operator in CBF calculations is commonly taken as a product of a state-dependent 
part and a scalar Jastrow-Feenberg part.

The main competitor to the many-body techniques outlined above 
is the coupled cluster method (CCM) \cite{Ray}.  It is based 
on describing the correlations in terms of exponentiated \emph{independent} excitations,  
which are parametrized as 
multiconfigurational creation operators with respect 
to some suitable reference state.  The CCM thereby completely avoids 
the complications arising from the overlapping products of correlation 
functions inherent in the Jastrow method.  A huge advantage is that it is now much easier 
to deal with state dependence, since the issue of non-commutativity of the correlation 
operators never arises.  A corresponding disadvantage is that it is much more 
difficult to deal with such extreme correlations as arise in hard-core systems in the 
second-quantized 
 representation that provides the natural framework 
for the CCM than it is in the first-quantized 
 representation in which the 
Jastrow and CBF methods are most naturally expressed.  

Both the CBF method and the CCM have been widely applied in many different areas of 
quantum many-body physics outside the field of nuclear physics to which both can trace 
their origins.  For example, the CCM has found many applications in quantum chemistry, 
where it 
is the method of first choice for very accurate descriptions 
of highly correlated atoms and molecules (see, e.g., Ref. \cite{Bartlett}).  
Recently, in this context, an attempt was made \cite{Nooijen} to extend the CCM 
to include 
more general excitations than those generated 
by exponentiated 
\emph{independent} two-particle/two-hole excitations.

This extension is tantamount to the 
use of state-dependent Jastrow wave functions, although the equations to be 
solved 
differ slightly from the usual variational 
ones, as discussed more fully below.  There are now claims in the literature that 
this method can give the \emph{exact} ground state wave function for systems interacting 
via pairwise forces \cite{Nooijen,Piecuch}.  We shall argue here that although the 
method can certainly be extremely accurate, it is \emph{not} in general exact.

Our fundamental concern is thus twofold: why is the Jastrow method so
accurate, and what limits its accuracy?  One obvious such limit is
clearly any constraint on the parametrisation of the correlation
functions. However, the key issue is that even if we use the most
general parametrisation possible, can the Jastrow method be exact?  In
view of our earlier discussion, it would clearly be surprising if it
were always exact, but in recent papers Nooijen \cite{Nooijen} and
Piecuch and his collaborators \cite{Piecuch} have claimed that this is
the case. Their basic idea is to work in a finite part of
occupation-number space, and obtain equations for the coefficients in
a second-quantized two-body operator that specifies the correlations.
Although there are an equal number of unknowns and equations, in
practice there seems to be a large degeneracy to the
solutions. Nevertheless, by tackling semi-realistic problems
they find such a high accuracy that the method
\emph{seems} exact. In this letter we analyse their method, and we
argue that such high accuracy is intrinsic to the Jastrow approach
underlying their calculations.  In particular, we analyse a simple
model with the aim of shedding some light on the issues raised above.

In the original Jastrow approach a correlated $N$-particle wave function
is decomposed as 
\begin{equation}
\Psi_{\alpha}(\vec{r}_{1},\ldots,\vec{r}_{N})=\bigl(\prod_{i<j}f(r_{ij})\bigr)\Phi_{\alpha}(\vec{r}_{1},\ldots,\vec{r}_{N})\quad,
\end{equation}
where $f(r_{ij})$ incorporates the effects of short-range
correlations between the particles, and $\Phi$ is a {}``simple''
wave function, typically Hartree or Hartree-Fock, which also incorporates
internal quantum numbers such as spin, etc., which we collectively
denote by the label {}``$\alpha$''.

We now rewrite the product of correlation functions in terms of a sum,
by means of an exponential representation, 
\begin{equation}
\prod_{i<j}f(r_{ij})=\exp\biggl(\sum_{i<j}u(r_{ij})\biggr).
\end{equation}
 The sum in the exponent can be recognized as a special 
case of a general two-body operator. If we look at 
state-dependent correlations which involve internal degrees of freedom 
as well as relative coordinates, the correlated wave function is 
characterized by a Jastrow \emph{operator}, 
\begin{equation}
\left|\Psi\right\rangle =\exp\bigl(\hat{T}_{2}\bigr)\left|\Phi\right\rangle ,\label{eq:correl2}
\end{equation}
 where the correlator $\hat{T}_{2}$ is a general two-body operator.

In the work of Nooijen \cite{Nooijen} it has been argued that
Eq.~(\ref{eq:correl2}) represents the exact ground state for
\emph{any} two-body Hamiltonian.  Although the examples
studied recently \cite{ex,Piecuch} provide numerical support for this
method, they are neither rigorous nor transparent.  This is partially
because they involve realistic or semi-realistic applications, where
the limitations and successes of the approach are less evident, and
partially because, as in most quantum-chemistry problems, the results
are largely perturbative, and thus do not provide a rigorous test of
the general method.

We can easily derive the form of Jastrow's method introduced by Nooijen
for use in a finite configuration space.
This starts from a general Jastrow operator in the form 
$
\hat{T}=\sum_{i}t_{i}\hat{O}_{i},\label{eq:defT}
$
where $\hat{O}_{i}$ is a complete set of two-body operators (we shall
also include one-body operators in our analysis, but the principle
remains the same). We label a complete normalised basis in the space by $\left|n\right\rangle $,
with $\left|0\right\rangle =\left|\Phi\right\rangle $. The ground-state
wave function is assumed to take the form 
\begin{equation}
\left|\Psi\right\rangle =e^{\hat{T}}\left|\Phi\right\rangle ,\quad\hat{H}\left|\Psi\right\rangle =E\left|\Psi\right\rangle \quad,
\end{equation}
where $\left|\Phi\right\rangle $ is a simple reference state. One
then evaluates the energy using a technique similar to that used in
the coupled cluster method, 
\begin{equation}
E=\left\langle \Phi\right|e^{-\hat T}\hat He^{\hat T}\left|\Phi\right\rangle \quad.\label{Eexpex}
\end{equation}
 We now use the fact that if $e^{\hat T}\left|0\right\rangle $ is an eigenstate
of the Hamiltonian, it must satisfy the equations \cite{H_Nouji}
\begin{equation}
0=\left\langle \Phi\right|e^{{\hat T}^{\dagger}}\hat O_{i}\hat  He^{\hat T}\left|\Phi\right\rangle 
-E\left\langle \Phi\right|e^{{\hat T}^{\dagger}}\hat O_{i}e^{\hat T}\left|\Phi\right\rangle \quad.\label{Nooijen1}
\end{equation}
Actually, very similar equations follow from the variational approach
to the Jastrow problems. By minimising the energy 
\begin{equation}
E=\frac{\left\langle \Phi\right|e^{{\hat T}^{\dagger}}\hat He^{\hat T}\left|\Phi\right\rangle }
{\left\langle \Phi\right|e^{{\hat T}^{\dagger}}e^{\hat T}\left|\Phi\right\rangle }\quad,\label{eq:Eexpecvar}
\end{equation}
and using hermiticity, we get Eq.~(\ref{Nooijen1}), apart from the fact
we must use Eq.~(\ref{eq:Eexpecvar}) for $E$ rather than the similarity
transformed expression in Eq.~(\ref{Eexpex}).

Finally, one can use the energy relation (\ref{Eexpex}) combined
with completeness to write the basic equations (\ref{Nooijen1}) in
the form derived in Nooijen's original work
\begin{equation}
\sum_{n\neq 0}\left\langle \Phi\right|e^{{\hat T}^{\dagger}}\hat O_{i}e^{\hat T}\left|n\right\rangle 
\left\langle n\right|e^{-\hat T}\hat He^{\hat T}\left|\Phi\right\rangle =0\quad.\label{eq:Nooijeqn}
\end{equation}

Since the number of equations in Eqs.~(\ref{Nooijen1},\ref{eq:Nooijeqn})
equals the number of unknowns in the operator $\hat{T}$, one might
expect that these equations have a solution. 
Since the equations are highly non-linear,
there
is no general proof of this assertion, and its validity must depend
on the nature of the Hamiltonian. One may argue that if $\left|\Phi\right\rangle $
is close to an eigenfunction of $\hat{H}$, a perturbative argument
will show that there may well be a solution to the problem.

One of the reasons to believe that the Jastrow method may  be exact 
is based on the technique of Euclidean filtering, as disccused 
in Refs.~\cite{Nooijen,Piecuch}.
This starts from the fact that the wave function
\begin{equation}
\left|\psi(t)\right\rangle ={e^{-\hat H t}
\left|\Phi\right\rangle }{/\left\langle \Phi\right|e^{-2\hat H t}
\left|\Phi\right\rangle ^{1/2}}\quad\label{eq:filter}
\end{equation}
approaches the exact one as $t\rightarrow\infty$, and that it is obviously a
two-body correlated wave function for all \emph{finite} $t$. If, in the
limit $t\rightarrow\infty$, the wave function
$\left|\psi(t)\right\rangle $ remains a two-body wave function, it
must be a solution to Eqs.~(\ref{Nooijen1}). For this
argument to be correct, it is a necessary and sufficient condition that
the set of two-body-correlated wave functions is complete.  The
non-trivial nature of such a statement can be seen from the fact that
we try to find a set of finite parameters in the operator $\hat T$ 
that can describe the same physics as obtained by
the limit $t\rightarrow\infty$ in Eq.~(\ref{eq:filter}).  Actually,
what we shall do below is provide an indirect proof that the exact
solution for a simple problem, which by the filtering argument must
also be the limit of $\left|\psi(t)\right\rangle $, is \emph{not}
two-body correlated.

A standard test-bed for many-body calculations is the Lipkin model.
This is a two-level model, where fermions can occupy either of two
levels. If we denote by $a_{\pm,i}^{\dagger}$ the creation operator of a fermion
in state $i$ in either the upper ($+$) or lower level ($-$), the
Hamiltonian of this model can be written in the form
\begin{equation}
\hat H=J_{0}+\lambda\left(J_{+}^{2}+J_{-}^{2}\right)\quad,
\end{equation}
where 
\begin{eqnarray}
J_{0}&=&\frac{1}{2}\sum_{i=1}^{\Omega}
\bigl( a_{+,i}^{\dagger}a_{+,i}-a_{-,i}^{\dagger}a_{-,i}\bigr)\quad,\nonumber\\
 J_{+}&=&\sum_{i=1}^{\Omega}a_{+,i}^{\dagger}a_{-,i},\quad J_{-}=J_{+}^{\dagger}\quad.
\end{eqnarray}
The fact that only $SU(2)$ generators appear means that we can diagonalise
states within different irreps of this algebra; here we shall concentrate
on the one with $J=\Omega/2$. Hence we shall only be considering
the states
\begin{equation}
\left|M=-J+n\right\rangle =\sqrt{\frac{(2J-n)!}{n!(2J)!}}(J_{+})^{n}\prod_{i=1}^{2J}a_{-,i}^{\dagger}\left|0\right\rangle \quad.\label{eq:basis}
\end{equation}

There are only a small number of two-body operators in the relevant
space, and most of these do not respect the $SU(2)$ dynamical symmetry
of the Hamiltonian. This leads to the only allowed two-body operators
being $J_{+}^{2}$, $J_{-}^{2}$, $J_{+}J_{-}$, $J_{-}J_{+}$, and
$J_{0}^{2}$. The last three operators are overcomplete.  Thus, when
acting in the basis of Eq.~(\ref{eq:basis},) all three of them
correspond to a quadratic function in $n$.  From the combination we
can then construct a constant, linear and quadratic piece. The
constant part is irrelevant for time-independent problems, and we only
need the linear and quadratic parts. These can be reached with the
operators $J_{0}^{2}$ and $J_{0}$ as well, which are what we shall
use. The single-particle operator $J_{0}$ does not seem to play a key
role. We see no \emph{a priori} reason why this operator is irrelevant
in the current calculation, apart from the fact that in the Jastrow
method one-body operators correspond to a general modification of the
single particle wave functions.

We now attempt to investigate perturbatively whether the
result from Nooijen's method agrees with the exact result. We write
\begin{equation}
t_{i}=\sum_{n=1}^{\infty}t_{i}^{(n)}\lambda^{n}\quad,
\end{equation}
and use as our reference state $\left|\Phi\right\rangle =\left|M=-J\right\rangle $,
the exact eigenstate for $\lambda=0$. There are three ways to solve
the problem.  Firstly, we can solve Eqs.~(\ref{Eexpex}) and (\ref{eq:Nooijeqn})
order by order in $\lambda$, which we shall refer to as {}``the
solution to Nooijen's equations''. Alternatively, we can either combine
Eq.~(\ref{eq:Nooijeqn}) with Eq.~(\ref{eq:Eexpecvar}) to {}``solve
the variational problem'', or we can write down the exact wave function
in Rayleigh-Schr\"{o}dinger perturbation theory, including all 
the arbitrary constants related
to its normalisation, and require that $e^{T}\left|0\right\rangle $
is equal to this wave function order by order in $\lambda$ (equality
of wave functions). All three approaches only involve matrix algebra,
which can easily be done with a computer-algebra package. The result
in each case is very instructive.

We look at these solutions as we vary $J$ from 1 to $8$, i.e., we
look at even particle numbers only. The number of wave functions mixing
in the ground state is easily seen to be $J+1$, so the number of
parameters in the wave function changes from larger than the number
of components, to much less than the number of components. Let us
first study the equality of the wave functions.

For $J=1,2,3$ we have no problems, and all indications are that the
wave function is exact. For all the other cases investigated we can
only have equality of the wave functions up to 7th order in $\lambda$.
If we insert this wave function in the energy expressions of Eqs.~(\ref{Eexpex},\ref{eq:Eexpecvar})
we find that the energy for the similarity transformed result is accurate
to the same order, whereas the variational estimate using the wave
function is actually accurate to 14th order 
\footnote{It is quite straightforward to understand this using a 
standard perturbative argument which can be used to  show that the
deviation of the energy is quadratic in any small fluctuation around the exact solution.}.

The solution to Nooijen's equations is equally illuminating. First
of all there appears to be an enormous degeneracy to these equations.
An obvious source for such degeneracy is in the choice of
normalisation of the wave function. Indeed, using intermediate
normalisation drastically reduces the number of independent variables,
but it doesn't fully lift the degeneracy.
We seem to be able to choose the coefficient of $J_{0}$ to equal zero,
without any loss of generality. We can also impose the requirements
that the coefficients of $J_{+}^{2}$ and $J_{-}^{2}$ are odd in
$\lambda$, and those of $J_{0}^{2}$ even. {[}The problems cited above
are totally independent of this choice, as has been checked for a
few representative cases.{]} An interesting observation is that the
these equations have two solutions at order $\lambda^{8}$,
one of which leads to an inconsistent set of equations at the next
order; the other continues. In the cases $J\leq3$ it continues forever, as far
as we can see, and when $J>3$ the solutions terminate at order $\lambda^{13}$ 
where we again get an inconsistent set of equations.
Surprisingly, this leads to a more accurate energy than above, up
to order $\lambda^{12}$ (which is almost, but not quite, as accurate as the
Jastrow method).  Note, however, that the lack of a solution to Nooijen's
equations gives rise to some problems. The deviations from zero may
actually be hard to see if $\lambda$ is small enough, which may be
an indication that for almost perturbative problems this is not such
a bad approach after all. 

Finally we can also solve the variational equations. It comes as no
surprise that, whatever the number of particles, we can solve this
problem. When $J>3$ the variational energy starts deviating from
the exact result at order $\lambda^{16}$, as expected.

From the simple model discussed here we can draw some conclusions on
the use of a Jastrow-like method in configuration space.  We have
disproven the idea that it can be generally exact, but to our surprise
it seems to be exact when the number of parameters in the correlation
operator is larger than the number of parameters in a general wave
function. Due to the non-linearity of the method, this is already a
highly nontrivial statement. When the method breaks down the wave
function is still correct to seventh order in perturbation theory,
which leads to a fourteenth-order accuracy for the Jastrow method.  If
we input this wave function into an estimate of the energy based on a
similarity transform of the Hamiltonian, we get a result that is only
correct to the same order as the wave function is correct. If we
determine the coefficients by what we called ``Nooijen's equations''
the accuracy of the energy can be improved to twelfth order - only a
little less accurate than the benchmark variational estimate!
Nonetheless, if one deals carefully with the enormous degeneracy,
numerical solution of the equations can be a valid approximation to a
many-body system. The problem is that there is only an approximate
solution to the equations, and so the method only works if the residual
interaction is small in one sense or another, as is often the case in
quantum-chemical calculations.

In conclusion, on the one hand we have the disappointing result that
the Jastrow wave function is not in general exact for  many-body
systems interacting via two-body forces, in contrast to recent
claims to the contrary.  However, the Jastrow wave functions
proposed in the new method may well be so accurate that in numerical
approximations they are sufficient to obtain energies of the accuracy
required, for example, in quantum chemistry calculations. 
It is surprising that the energy estinate of  Eq.~(\ref{Eexpex})
is \emph{almost} as accurate as the optimal variational estimate of
Eq.~(\ref{eq:Eexpecvar}). We see no intrinsic reason for this.
Nevertheless, even when the
parameters are derived from a solution to Nooijen's equations,
its is more acuurate to use the variational estimate Eq.~(\ref{eq:Eexpecvar}).


\begin{thebibliography}{10}
\bibitem{Feenberg}J. W. Clark and E. Feenberg, Phys. Rev. \textbf{113}, 388 (1959);
E. Feenberg, \emph{Theory of Quantum Fluids} (Academic press, New
York, 1969). 
\bibitem{Jastrow} R. Jastrow, Phys. Rev. \textbf{98}, 1479 (1955). 
\bibitem{VMC}D. M. Ceperley and M. H. Kalos, in \emph{Monte Carlo Methods in Statistical
Physics}, ed.~K. Binder (Springer-Verlag, Berlin, 1979). 
\bibitem{Gaudoin} R. Gaudoin
\emph{et al.}, 
Phys. Rev. B \textbf{63}, 115115 (2001).
\bibitem{FHNC}J. W. Clark, in \emph{Progress in Particle and Nuclear Physics} \textbf{2},
ed.~D. H. Wilkinson (Pergamon, Oxford, 1979). 
\bibitem{FS}K. E. Schmidt and S. Fantoni, Phys. Lett. B \textbf{446}, 99 (1999). 
\bibitem{CBF}S. Fantoni and V. R. Pandharipande, Phys. Rev. C \textbf{37}, 1697
(1988). 
\bibitem{Ray}R. F. Bishop, Theor. Chim. Acta \textbf{80}, 95 (1991); in: {}``\emph{Microscopic
Quantum Many-Body Theories and Their Application},'' J. Navarro and
A. Polls (eds.), Lecture Notes in Physics \textbf{510},
(Springer, Berlin, 1998), p. 1.
\bibitem{Bartlett} R. J. Bartlett, J. Phys. Chem. \textbf{93}, 1697 (1989). 
\bibitem{Nooijen}M. Nooijen, Phys. Rev. Lett. \textbf{84}, 2108 (2000).
\bibitem{Piecuch}P. Piecuch
\emph{et al.},
Phys. Rev. Lett.
\textbf{90}, 113001 (2003).
\bibitem{ex}T. Van Voorhis and M. Head-Gordon, J. Chem. Phys. \textbf{115}, 5033
(2001); M. Nooijen and V. Lotrich, J. Chem. Phys. \textbf{113}, 4549
(2000) 
\bibitem{H_Nouji}H. Nakatsuji, Phys. Rev. A \textbf{14}, 41 (1976); J. Chem. Phys.
\textbf{115}, 2465 (2001); J. Chem. Phys. \textbf{116}, 1811 (2002). 
\end{thebibliography}
\end{document}